# A Theory Building Study of Enterprise Architecture Practices and Benefits

**Ralph Foorthuis**
UWV, CIO Office and Data Services
La Guardiaweg 116
1040 HG Amsterdam, The Netherlands
ralph.foorthuis@uwv.nl

**Marlies van Steenbergen**
Sogeti Netherlands
Postbus 76
4130 EB Vianen, The Netherlands
marlies.van.steenbergen@sogeti.nl

**Sjaak Brinkkemper**
Utrecht University, Information and Computing
Sciences, Princetonplein 5
3584 CH Utrecht, The Netherlands
s.brinkkemper@uu.nl

**Wiel Bruls**
IBM Netherlands
David Ricardostraat 2-4
1066 JS Amsterdam, The Netherlands
wiel_bruls@nl.ibm.com

**Abstract.** *Academics and practitioners have made various claims regarding the benefits that Enterprise Architecture (EA) delivers for both individual projects and the organization as a whole. At the same time, there is a lack of explanatory theory regarding how EA delivers these benefits. Moreover, EA practices and benefits have not been extensively investigated by empirical research, with especially quantitative studies on the topic being few and far between. This paper therefore presents the statistical findings of a theory-building survey study (n=293). The resulting PLS model is a synthesis of current implicit and fragmented theory, and shows how EA practices and intermediate benefits jointly work to help the organization reap benefits for both the organization and its projects. The model shows that EA and EA practices do not deliver benefits directly, but operate through intermediate results, most notably compliance with EA and architectural insight. Furthermore, the research identifies the EA practices that have a major impact on these results, the most important being compliance assessments, management propagation of EA, and different types of knowledge exchange. The results also demonstrate that projects play an important role in obtaining benefits from EA, but that they generally benefit less than the organization as a whole.*

**Keywords:** Enterprise Architecture, Architectural Insight, Compliance, Projects, Organizational Performance, Benefits, Partial Least Squares, Formative Measurement, Question-driven Research, Exploratory Study

# A THEORY BUILDING STUDY OF ENTERPRISE ARCHITECTURE PRACTICES AND BENEFITS

## 1. INTRODUCTION

By providing holistic overviews and norms, Enterprise Architecture (EA) aims to achieve coherent and goal-oriented organizational processes, structures, information provision and technology (cf. Boh & Yellin 2007; Richardson et al. 1990; Ross et al. 2006; Tamm et al. 2011; The Open Group 2009; Wagter et al. 2005). Academics and practitioners have made various claims regarding the application and effectiveness of EA (e.g. Bucher et al. 2006; Gregor et al. 2007; Mulholland & Macaulay 2006; Ross et al. 2006; Tamm et al. 2011). At the level of the entire organization, for example, benefits in the reduction of complexity and realization of business/IT alignment have been claimed. At the level of individual projects, costs and risks are said to be reduced when using EA.

However, Enterprise Architecture research has stumbled upon three fundamental problems. First, there is a lack of explanatory theory in the field of EA (Boucharas et al. 2010; Lange et al. 2012b; Radeke 2010; Tamm et al. 2011). Many publications focus on how to create an EA or EA function but neglect to investigate the arguably most important topic, namely how a given EA can realize benefits. The publications that do present such explanatory insights tend to deliver fragmented contributions because they often focus on a single aspect, such as alignment (e.g. Bradley et al. 2011; Gregor et al. 2007; Tiwana & Konsynski 2010). Moreover, many publications deliver only implicit views on how EA yields benefits, as they present a non-explanatory gray box model or no explicit model at all (e.g. Boh & Yellin 2007; Niemi 2006; Slot et al. 2009; Schmidt & Buxmann 2011). The result is thus a paucity of clear and explicit theory on *how* EA brings value to the organization. This is problematic because explanatory studies identifying internal mechanisms for achieving the organization's end goals are crucial in IS theory, not in the last place for understanding which success factors are required (Schryen 2013). A second acknowledged problem in the field of EA is the lack of empirical research and, more specifically, quantitative studies on how EA delivers benefits (Boh & Yellin 2007; Bradley et al. 2011; Kappelman et al. 2008; Niemi 2006; Radeke 2010; Tamm et al. 2011). Third, although the important role of projects and the issue of whether they correctly use (comply with) EA has been acknowledged in the literature (Ross et al. 2006; Tamm et al. 2011; Wagter et al. 2005; Ylimäki et al. 2007), it has not been a topic of much research and is usually not well-integrated into EA theory (Bandara et al. 2007; Foorthuis 2012; Kaisler et al. 2005).

To address these three problems and give renewed impetus to EA-research in order to advance theory, this study investigates how EA delivers value to organizations. We start with a general theoretical framework that offers explanatory potential and is capable of accommodating the claims regarding EA practices and benefits. In interaction with existing and often implicit theory on EA, our exploratory study advances this framework into a synthesized overall explanatory model of EA practices and benefits. To ensure that our modeling efforts not only take into account existing views but are also consistent with the empirical world, we have conducted a survey study ($n$=293). Due to the theory-building nature of our research, this paper does not follow the logic of a standard confirmatory study. More specifically, explanatory EA-theory and hypotheses that are sufficiently sharp for testing are not the starting point of our research; rather they are the end result. Our final model is the product of an iterative research process, with empirical data, fragmented and implicit theory on EA, and the many research discussions serving as input. The resulting model makes clear how EA practices and intermediate outcomes contribute to achieving goals at both the organizational and the project level. The research also identifies the EA practices that have a major impact on the effectiveness of EA.

The overall research question of this paper is:

> *How does EA deliver benefits, and what are the most effective EA practices for working with EA?*

This paper proceeds as follows. In section 2 we present our research approach. In section 3 we present an overview of relevant literature, as well as the explanatory model that forms our core contribution. In section 4 we discuss the statistical analysis of our empirical data. In section 5 we present the conclusions.

## 2. AN EXPLORATORY RESEARCH APPROACH

We have employed an *exploratory* or *question-driven* research approach, in which immature ideas, curiosity and empirical data are iteratively put to use to create and advance scientific models (Firestein 2012; Glass 2010; Jewett 2005; cf. Wold 1980). Proponents of this research framework criticize hypothesis-driven studies (aka confirmatory or test-based studies) on several grounds. First, hypothesis-driven research makes the researcher prone to subjectivity and even bias, because he enthusiastically wants to confirm his own ideas (Greenwald et al. 1986; Glass 2010; Jewett 2005; Firestein 2012). Second, although empirical studies are usually presented in a confirmatory or test-based fashion, they are in fact often question-driven, iterative and exploratory in nature (Feelders 2002; Glass 2010; Henseler et al. 2014; Rigdon 2012; Sarstedt et al. 2014). In that sense, our study is thus not so fundamentally different from mainstream research, except for the fact that we openly acknowledge its iterative nature. Fortunately, it is increasingly recognized in the fields of IS, marketing and management (Chin 2010; Henseler et al. 2014; Ko & Osei-Bryson 2008; Padmanabhan et al. 2006; Sarstedt et al. 2014; Shmueli & Koppius 2011; Bidan et al. 2012) as well as in statistics and econometrics (Feelders 2002; Shmueli 2010; Wold 1980) that, while testing remains necessary, such exploratory and iterative statistical research can play an important role in theory generation.

Our study should indeed be considered exploratory, as "few researchers have analyzed and generalized EA [management] phenomena, whereas the share of contributions to explanation and prediction is negligibly small" (Radeke 2010, p. 1; cf. Tamm et al. 2011). Indeed, our final model, based on survey data and the fragmented literature, has not been presented in other publications before. We also started our study with a rather general framework (Figure 1) that required transformation to an explanatory model. Moreover, we based our constructs partly on theory (which inspired the inclusion of new model constructs) and partly on empirical data (which banned constructs with insignificant paths from the final model due to lack of nomological validity). Finally, the immaturity of theory made it inappropriate to simply assume linear effects. Therefore, we had no *a priori* expectations regarding the (non)linearity of relationships and thus had to investigate several possibilities.

Our primary statistical method is PLS, since this variance-based SEM-method is well-suited for exploratory studies (Chin 2010; Gefen et al. 2000; Hair et al. 2010, 2011; Henseler et al. 2014; Urbach & Ahlemann 2010; Wetzels et al. 2009). One reason for this is that PLS' insensitivity to indeterminacy problems allows the successive building of models in an iterative fashion (as opposed to covariance-based SEM). Furthermore, as a limited information technique, wrongly specified and tentative sub-parts of the model do not ripple through the entire model estimation process. Finally, PLS offers several overall model evaluation criteria when developing theory. In addition to the support for theory building, PLS offers several other advantages. First, this technique not only allows path modeling with aggregated constructs, but also handles both formative and reflective types (Gefen et al. 2000; Haenlein & Kaplan 2004; Hair et al. 2010; Urbach & Ahlemann 2010; Wetzels et al. 2009). Second, PLS is nonparametric in nature (Chin 2010; Haenlein & Kaplan 2004; Hair et al. 2010), which is an advantage since our variables have been measured mostly at the ordinal level and we have no a priori distributional assumptions. Third, PLS can easily handle both our single-item variables and multi-item constructs (Hair et al. 2010). We refer the reader to section 4.1 for the details of our research design.

Note that *exploratory* refers to our research approach (see above), while *explanatory* characterizes the nature of our substantive contributions. Regarding the latter, our study aims to deliver an EP (type IV) theory, i.e. theory for explaining and predicting (Gregor 2006). Such a theory focuses amongst others on how the world works (explaining) and is able to generate testable hypotheses for future confirmatory research (predicting). As explaining and causation are "intimately linked" and explanation implies "laws" as well as a "communicative process" (ibid.), we will present not only a statistical path model but also a description and interpretation regarding causal relationships.

## 3. LITERATURE OVERVIEW AND THEORY

In this section we discuss the extant literature and present our final model. We used a wide array of both academic and practitioner publications for our enquiry. This included, but was not limited to, a search in the journals MISQ, ISR, JAIS, EJIS, JIT and JSIS for the keywords "architecture", "architect" and/or "architectural" in the articles' titles from 2006 to June 2014. Section 3.1 presents the general framework that formed the theoretical basis at the very beginning of our exploratory study. Sections 3.2 and 3.3 present the insights from the literature that served as input for transforming the general framework into an empirically-based explanatory model. In this regard, section 3.2 identifies the practices for and benefits of working with EA. Section 3.3 presents explanatory insights that iteratively inspired our model building efforts throughout our research project. Section 3.4 presents our final model.

### 3.1. GENERAL THEORETICAL FRAMEWORK AND DEFINITIONS

Figure 1 presents the high-level theoretical framework that formed the starting point for our exploratory study. This overall framework shows how the concepts in our study on the application and effectiveness of EA are interrelated at the highest level of abstraction. Note that the included practices and benefits are merely examples to clarify the framework (see section 3.2 for a full overview). We define *Enterprise Architecture* as the set of high-level views and norms that guide the coherent design and implementation of processes, organizational structures, information provision and technology within an organization (Foorthuis and Brinkkemper 2008; Lankhorst et al. 2005; Richardson et al. 1990; Wagter et al. 2005). EA norms are prescriptions such as principles, models and policy statements (e.g. "Use technology *x* for workflow management"). An organization's EA in itself, as a set of documents, offers no value if it is not used in practice (Persson & Stirna 2001; Wagter et al. 2005). Therefore, several practices (techniques) should be employed to work with and stimulate the use of EA (Boh & Yellin 2007; Goodhue et al. 1992; Tamm et al. 2011). For example, knowledge and practical assistance can be offered to projects when they apply the EA norms and projects can be assessed on conformance. A well-balanced combination of practices constitutes an *EA approach*. Employment of such an approach should lead to the effective use of and project compliance with EA (Bandara et al. 2007; Kaisler et al. 2005; Ylimäki et al. 2007). *Compliance* (and its synonym *conformance*) in this context represents the *correct use of EA*. More specifically, it is the extent to which there exists accordance between behavior or products on the one side and predefined explicit EA norms on the other (Foorthuis et al. 2012; Kim 2007). The correct use of the organization's EA is necessary for reaping the aforementioned architectural *benefits* (Lange et al. 2012a; Wagter et al. 2005; cf. DeLone & McLean 2003). These purported benefits are multifold. For the *organization as a whole*, for example, the organization becomes able to align business and IT, respond agile, and – instead of local optimizations – implement a coherent enterprise-wide strategy. In terms of *projects*, for example, desired quality and functionality can be delivered, while costs and complexity are said to be controlled. Although often left implicit in the literature, we expect intermediate benefits to affect end benefits (Boucharas et al. 2010; Radeke 2010; Tamm et al., 2011). Several examples can be provided in this context. First, the insights into the organization offered by EA is often seen as one of its most fundamental added values (Bernard 2012; Lankhorst et al. 2005; Tamm et al. 2011). Insight-yielding benefits should thus have an effect on benefits such as complexity management (Gregor et al. 2007; Lankhorst et al. 2005). Second, we can expect project results to have an effect on enterprise-wide benefits, since implementation projects are needed to achieve, for example, agility at the organizational level (Ross et al. 2006). Third, EA practices themselves can yield benefits, such as when improved organizational insight is the direct result of organized knowledge exchanges. In addition to the above, the effects of techniques and conformance might be affected by several *contextual factors* (Schryen 2013), such as the economic sector, organizational size and EA focus (on business, IT or both).

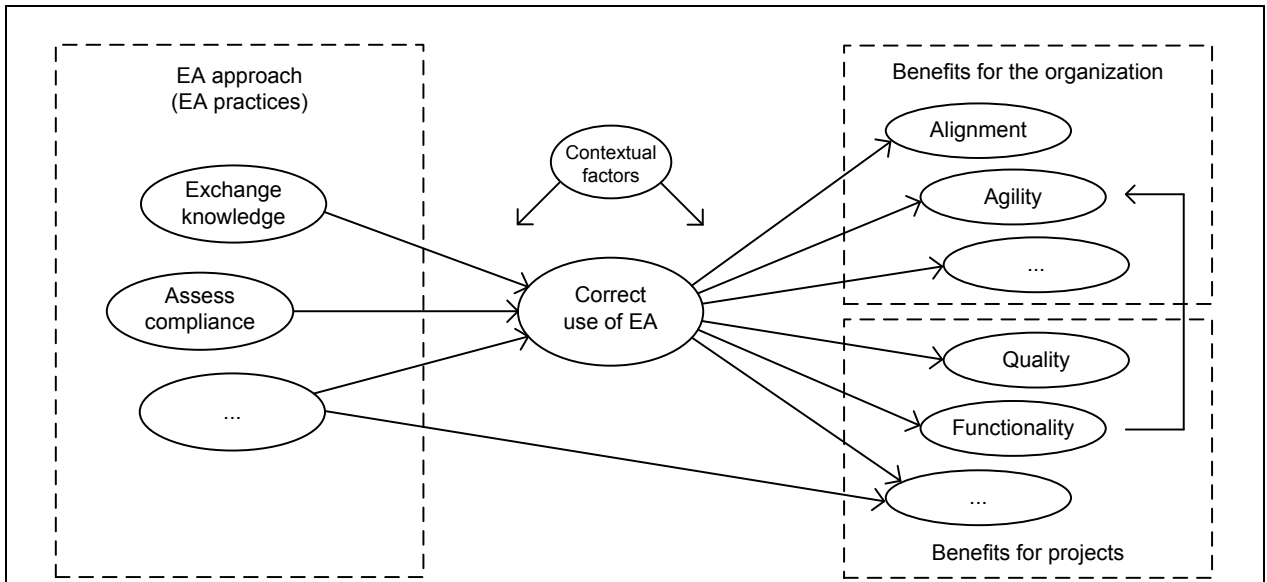

**Figure 1. General theoretical framework for EA practices and benefits (includes examples)**

The framework of Figure 1 has been used to structure our research. As is often the case in IS research, this model is a 'gray box' (Schryen 2013) since it does not explain *how* EA delivers value. Therefore, advancing this general framework and the fragmented insights from the literature into an explanatory, more sophisticated and empirically supported model of EA practices and benefits is the primary aim of this research paper.

## 3.2. OVERVIEW OF CLAIMED EA PRACTICES AND BENEFITS

In this section we discuss the individual *practices* (techniques) for and *benefits* of working with EA in more detail. We start by presenting the practices. To be able to reap the benefits of EA, it is important that an organization's EA is actually used and complied with (Boh & Yellin 2007; Goodhue et al. 1992; Ylimäki et al. 2007; Tamm et al. 2011). The organization therefore needs practices that can be employed to work with EA in the way intended by the architects. Due to our research question, these techniques are selected for their potential to stimulate correct use of a given EA (practices for creating an EA or EA function are thus less relevant here). For later reference, each practice is coded in parentheses with a capital T (e.g. T1).

**Ensure management involvement in EA.** EA should enable the achievement of strategic business goals (Morganwalp & Sage 2004; Obitz & Babu 2009). In this context it may be important for management to formally approve the EA (T1) (Lange et al. 2012b; Van Steenbergen et al. 2010). Management should also ensure that the choices in the EA are explicitly linked to the strategic business goals (T2) (Van Steenbergen et al. 2010). Furthermore, it is necessary for management to actively propagate the organization's EA, e.g. by emphasizing the importance of EA and its norms (T3) (Boh & Yellin 2007; Lange et al. 2012b; Radeke 2010).

**Assess EA conformance.** Projects should be monitored on compliance with the EA's constraints, standards and other norms (T4) in order for the organization to be able to take corrective action (Boh & Yellin 2007; Foorthuis et al. 2012). This often takes the form of a formal review and approval process (Schmidt & Buxmann 2011).

**Create an active community for EA knowledge exchange.** The division of architectural domains with their own domain architects, which is often felt necessary in large organizations, carries the inherent risk of fragmentation and misalignment. An active community of EA practice should enable knowledge integration (Andersson et al. 2008; Van Steenbergen 2011). This manifests itself in organized knowledge exchanges between architects (T5) and between architects and project members (T6). Moreover, it is regularly stressed that architects should be actively involved in projects, as they can assist the projects in defining the solution and applying EA norms (T7). For example, they can actively participate in the project (Lange et al. 2012a; Slot et al. 2009; Wagter et al. 2005).

**Leverage the value of project artifacts.** A Project Start Architecture (PSA) is a document created at the start of a project. This deliverable inherits and translates the EA's prescriptions – such as rules, guidelines and models – to the specific project situation (Wagter et al. 2005). A PSA can be regarded as a specific form of what The Open Group (2009) refers to as an architecture contract. It describes the tangible constraints within which the project must operate. Using a PSA (T8) can therefore encourage the project to comply with the EA's norms (Wagter et al. 2005). In fact, document templates in general (T9) can be employed to increase architectural conformance and insight for two reasons. First, they can be used to increase knowledge integration between architects (Van Steenbergen 2011). Second, templates can be used to provide projects with pre-defined EA content, and provide the authors with instructions on how to conform (Foorthuis & Brinkkemper 2008; Wagter et al. 2005).

**Use compensation or sanctioning for stimulating conformance.** Incentives and disincentives could increase the willingness to think enterprise-wide and to comply with EA (T10). For example, the IT-costs associated with conformance might be compensated for by the EA program (Foorthuis et al. 2012; cf. Lange et al. 2012a) or incentives for IT professionals can be based on enterprise-wide performance (Ross et al. 2006).

In the remainder of this section we present an overview of claimed *benefits* of working with EA. These benefits are indicated with a capital B (e.g. B1). Several types of benefits can be identified. First, the *organization as a whole* should profit from EA by being able to gain multiple benefits:

**EA enables management to achieve key business goals.** First, EA is said to enable management to pursue a coherent strategy that is optimal for the entire enterprise, instead of local optimizations (B1). Individual domains and departments may strive to pursue local interests (Malloy 2003). However, the firm as a whole will not benefit from conflicting goals and an EA can provide the required holistic view of the enterprise to balance different interests and solutions (Lankhorst et al. 2005; Richardson et al. 1990). In addition, by taking this holistic and multi-layered view and providing insights, Enterprise Architecture is a valuable instrument for aligning IT and the business processes it supports (B2) (Bradley et al. 2011; Bucher et al. 2006; Gregor et al. 2007; Lankhorst et al. 2005).

**EA enables management of organizational complexity.** Complexity can be managed (B4) by improved insights, a modular approach – which distinguishes between parts of a system and their relationships – and architectural modeling languages (Lankhorst et al. 2005; Tamm et al. 2011; Versteeg & Bouwman 2006). Furthermore, implementing standardized and automated processes should result in less complex technology environments (Radeke 2010; Ross et al. 2006; Tamm et al. 2011).

**EA facilitates the integration, standardization and deduplication of processes and systems.** Years of organic growth have often led to various 'silos' or 'stovepipes', which do not leverage the potential of related or similar processes and systems. The high-level overviews of an EA provide insights into the organization's processes, business structures, information and IT-systems. This enables the enterprise to identify processes that could be integrated (for sharing valuable information), standardized (for supporting similar processes by the same systems) or even cut out (for decreasing redundancy) (B5) (Bidan et al. 2012; Morganwalp & Sage 2004; Kim et al. 2007; Niemi 2006; Ross et al. 2006). As a result of this and of the aforementioned complexity management, costs can be controlled (B6) (Lange et al. 2012b; Schmidt & Buxmann 2011).

**EA enables the enterprise to deal with its environment effectively.** Markets and businesses change ever more rapidly nowadays, and business processes and systems are highly interdependent. This poses problems for IT, as software has to be updated or replaced sooner whilst simultaneously being part of an increasingly complex environment. EA can increase the agility of the enterprise (B7) by automating the core business processes and by identifying the most promising projects (Bradley et al. 2011; Lange et al. 2012a; Ross et al. 2006). This results not only in additional resources, but also in valuable information, which can be used in innovative activities. In addition, by focusing on the contextual relationships, identifying key stakeholders and managing outsourcing arrangements, an EA can facilitate co-operation with other organizations (B8) (Bradley et al. 2011; Jonkers et al. 2006; Morganwalp & Sage 2004; Ross et al. 2006; Tamm et al. 2011).

A second type of key EA benefit is *gaining insights and understanding* regarding the IST and SOLL situations:

**EA yields insights into the current situation.** Architecture can provide insights into complex situations and problems (B3). Insights can be gained by overviews in terms of visual models, aspect areas (e.g. business, information, applications and infrastructure), compositionality and levels of abstraction (Capgemini 2007; Lankhorst et al. 2005; Raadt et al. 2004; Tamm et al. 2011). These insights can be used to improve and speed up decision making (Bernard 2012; Tamm et al. 2011). Moreover, these overviews and clear definitions of concepts provide organizational members with a shared frame of reference to communicate effectively with each other (B10) (Armour et al. 1999; Bernus 2003; Kappelman et al. 2008; Raadt et al. 2004). This should resolve conflicts, encourage cooperation and prevent redundancy (Tamm et al. 2011).

**EA yields insights into the future situation.** The techniques for presenting insights into the current situation can also be used to present a consistent and coherent overview of the desired state of affairs (Gregor et al. 2007; Lankhorst et al. 2005; Radeke 2010; Ross et al. 2006). Moreover, architectural norms (e.g. principles and models) provide tangible prescriptions to guide change initiatives (Richardson et al. 1990; Wagter et al. 2005). EA thus provides a clear description of the desired future situation (B9).

A third type of purported EA benefit concerns the *increased performance of individual projects that conform to EA*:

**Working with EA reduces project costs and project duration.** Projects can be expected to save resources (B12) and time (B13), since EA norms guide development work. EA and domain level decisions are a given starting point and do not have to be discussed inside the project (Capgemini 2007; Mulholland & Macaulay 2006; Pulkkinen & Hirvonen 2005; Wagter et al. 2005). A project can thus quickly focus its attention on designing and developing the details of the solution.

**Working with EA reduces project risk and improves project success.** Although often discussed only superficially, EA is said to identify and mitigate project risks (B14). The argument usually put forward is that EA models – with their views on platforms, applications, processes and relationships with other projects – provide an insight into project risks (Bucher et al. 2006; Capgemini 2007; Lange et al. 2012b), allowing for timely risk prevention tactics. In addition, projects that conform to EA can benefit from the fact that issues at the enterprise-level have already been solved in the EA, thus mitigating risk and increasing the probability of success, instead of building on sand (Capgemini 2007; Mulholland & Macaulay 2006; Pulkkinen & Hirvonen 2005). Because of the EA's standard framework and holistic overviews, projects can also be managed better (Bradley et al. 2011; Lange et al. 2012a). On a similar note, EA can be used to align the project with its context, resulting in high quality (B15) and relevant functionality (B16).

**Working with EA enables projects to manage complexity.** EA is said to enable projects to deal with complexity (B17) (Capgemini 2007; Lange et al. 2012a). Analogous to controlling complexity at the organizational level, EA facilitates management of project complexity by using aspect areas, levels of abstraction, a modular approach, up-front decision making, and by standardized services, processes and systems (ibid.; Lankhorst et al. 2005; Ross et al. 2006). This should simplify project tasks, especially since certain issues should already have been investigated and resolved by the EA function.

**Working with EA speeds up the initialization of a project.** An EA provides models of the enterprise, which help to specify the project scope and avoid redundant development activities (Bucher et al. 2006). Furthermore, several decisions have been made up-front and can be readily leveraged, for example by using a PSA (Wagter et al. 2005). Therefore, projects that have to conform to EA are expected to get initialized relatively fast (B18).

### 3.3. TOWARDS CONSTRUCTS AND RELATIONSHIPS

As shown in the Introduction, the EA literature is quite fragmented (individual studies focusing on a single EA topic), often implicit (no explicit causal models)[1] and usually not based on empirical data. For this reason our study aims to present a synthesized and statistically evaluated model that explicitly shows how EA delivers value. This section therefore gives an overview of the literature that, in interaction with our statistical analysis and research discussions, contributed crucial elements to our model. It shows how the extant literature inspired and theoretically supports the constructs and relationships of this study's primary model, which is presented in section 3.4. Italic text refers to the constructs of our model described in section 3.4.

It is widely accepted that *compliance* is important for obtaining value from EA (e.g. Bandara et al. 2007; Boh & Yellin 2007; Foorthuis et al. 2012; Kaisler et al. 2005; Schmidt & Buxmann 2011; Ylimäki et al. 2007). Although this is almost always neglected in explanatory research (i.e. no explicit construct), Boh and Yellin do present a causal model in which conformance with EA takes on a central and explicit role. In their view, compliance should be achieved by various governance mechanisms, such as liaison roles and monitoring compliance. Compliance, in its turn, results in various benefits, such as reduced heterogeneity and complexity, and increased integration of applications and data. This is similar to our study because we view compliance as a central mediating variable, which is influenced by an EA approach featuring compliance assessments, assistance and organized knowledge exchanges between various roles. Like Boh and Yellin, we see the concept of compliance as representing the (correct) use of EA. However, while these authors view compliance as the only mediator, we regard it as merely one of several important constructs in an extensive causal mechanism. Furthermore, we focus more on the role of projects in this context.

Many authors also see *architectural insight* as an intermediate but fundamental EA outcome (e.g. Bernus 2003; Lankhorst et al. 2005; Kappelman et al. 2008; Raadt et al. 2004; Tamm et al. 2011). In his overview, Radeke (2010) mentions "transparency" as an important EA benefit, and Ross et al. (2006) state that EA is a tool for communicating the company's direction. According to Bernard (2012) standardized analysis methods, shared reference information, improved understanding of processes and an integrated view of the enterprise are all crucial insights. The improved decision making afforded by them results in various intermediate benefits, such as resource deduplication, integration, more effective (project) management and improved business processes. These, in turn, result in reduced costs, quicker implementation and better overall performance. This is highly similar to our model, although projects and compliance play a more salient role in our study.

The literature also recognizes that achieving end goals is neither the direct result of EA nor of architectural insight. Although often intangible in nature, it is crucial to acknowledge the intermediate role of these internal capabilities in IS research (Schryen 2013). The importance of such mediating and organization-specific capabilities is also acknowledged in resource-based theory, both in general (Barney 1991; Wade & Hulland 2004; Aral & Weill 2007) and in the context of EA (Lux et al. 2010). Tiwana & Konsynski (2010) and Van Steenbergen (2011) explicitly acknowledge the importance of intermediate EA-induced outcomes that should result in the achievement of end goals. Furthermore, Ross et al. (2006) think in terms of a foundation for execution, which implements the core IT systems and business processes that support the organization's strategy. In their view this equates to effective levels of process digitization, data integration and accessibility, standardization and flexibility, and shared IT services. With this foundation for execution, the organization is able to realize its business strategy and achieve its end goals. These ideas are consistent with the *EA-induced capabilities* in our model, which represent these foundational intermediate outcomes.

---

[1] Tamm et al. (2011) have made a worthwhile effort to render some of these implicit views explicit.

Furthermore, all theories mentioned above devote attention to what we will refer to as the *EA approach*, which is related to concepts such as EA management, governance and architecture practices. Likewise, all theories explicitly refer to different types of end goals or performance. While many publications on EA focus mainly on enterprise-wide *organizational performance*, multiple publications also acknowledge the more localized benefits of *project performance*. Our model therefore includes both types of benefits.

Some may argue that the DeLone and McLean model of IS success (DeLone & McLean 2003) and its application on EA (Lange et al. 2012a) already constitute sufficiently mature theory. However, this is not the case for our purposes. First, these DMSM models have a fundamentally different focus, shedding light on what characteristics stimulate further usage of a system (or EA) rather than on how such usage is able to realize benefits. Use in these models directly leads to end benefits, which thus constitutes a gray box model instead of a satisfactory explanation of *how* usage delivers value. They explain *use* of systems, not their organizational *benefits*. Second, the practices stimulating the use of a system (or EA), such as knowledge exchanges, are also absent from these models. Similar to DMSM models, however, we consider "use" a very important element. While this construct has been criticized for being vague (cf. DeLone & McLean 2003; Lange et al. 2012a), our interpretation of it is very specific, namely projects complying with EA.

Our model is also consistent with general IS theory on business value (Schryen 2013). In this theory the impact of IS investments (the EA and EA practices) on performance (e.g. agility) is mediated through the intermediate benefits of process performance (e.g. architectural insight) and affected by contextual factors (e.g. economic sector). In this context it is important to disaggregate IS business value (ibid.), which we have done not only by differentiating between different EA benefit constructs, but also by measuring them formatively and thus acknowledging their sub-benefits.

### 3.4. The Explanatory Model

This section presents the explanatory model (Figure 2), which is the core contribution of this study. We have incorporated six EA-related formative constructs into our model. Sections 3.2, 3.3 and 4.3 provide theoretical and empirical support for these constructs, while the current section describes their meaning. We define the *EA Approach* as the set of practices that the organization employs for working with EA and for having projects comply with architectural norms. The EA Approach includes practices such as compliance assessments, knowledge exchanges, formal approval of EA and management support. See Table 5 for a concise overview of techniques comprising this construct (items T1 to T10). *Project Compliance with EA* represents the degree to which the organization's projects adhere to the EA principles, models and other norms (item O1). Therefore, this variable represents the correct application of the EA and its norms. *Architectural Insight* is also specific to EA and represents the degree of EA-induced organizational knowledge and understanding regarding the IST and SOLL situations. Rather than simply having knowledge of the EA norms, architectural insight refers to a fundamental, practical and shared understanding of the nature and relations of relevant systems and processes, organizational goals, constraints and complexity. It therefore comprises knowledge of the complexity of the organization (B3), a clear image of the desired future situation (B9), communicating effectively (B10) and a shared interpretation of the EA norms (O2).

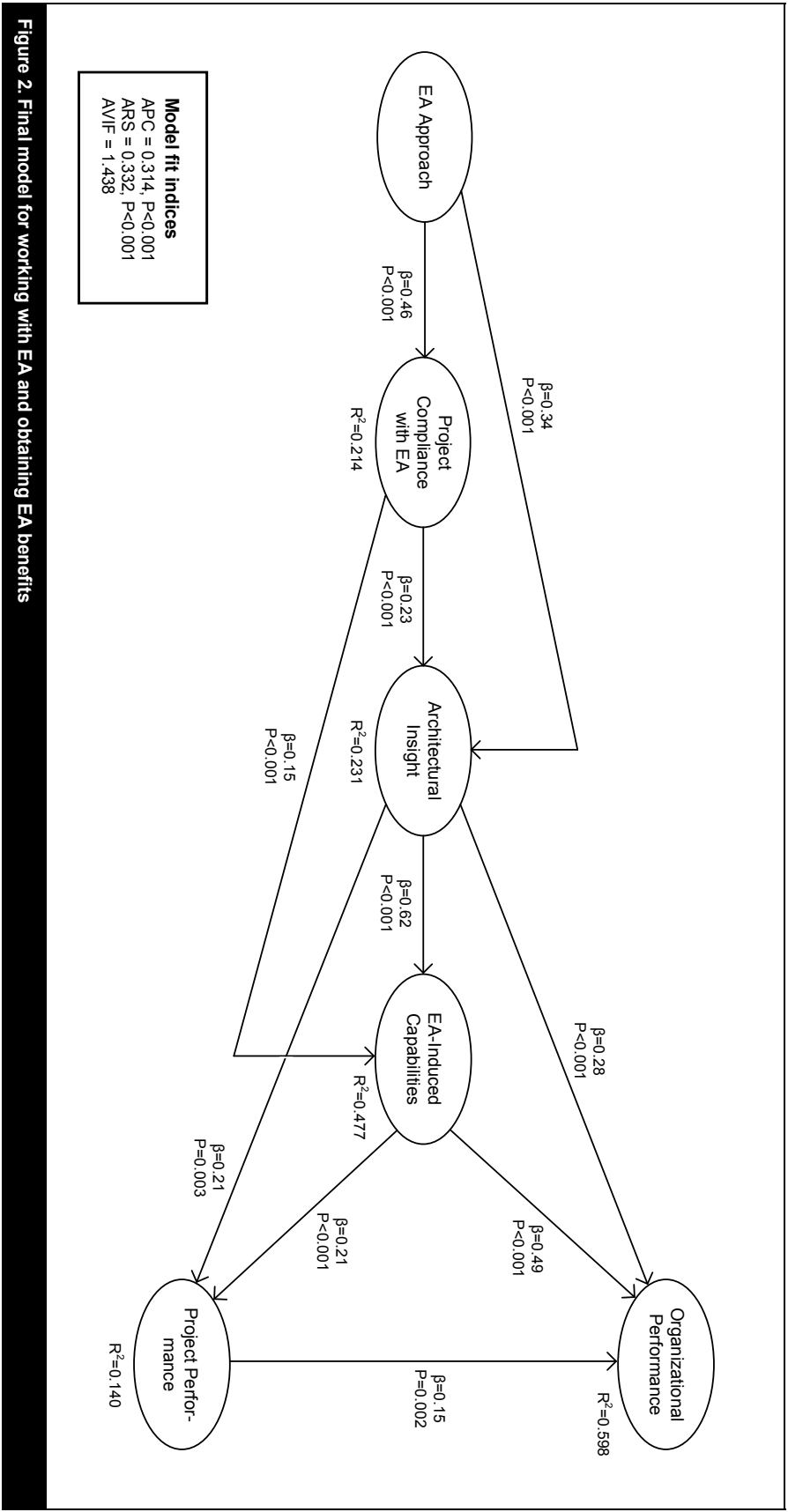

**Figure 2. Final model for working with EA and obtaining EA benefits**

**Model fit indices**
APC = 0.314, P<0.001
ARS = 0.332, P<0.001
AVIF = 1.438

The construct *EA-Induced Capabilities* represents the organization-specific expertise and modes of working required to implement foundational EA-related benefits that are necessary for obtaining typical EA end goals. These are typically EA benefits that are valuable in themselves, but should be regarded as intermediate benefits from an end goal perspective. This construct therefore includes the often-mentioned abilities to control complexity (B4), integrate, standardize and deduplicate processes and systems (B5), align business and IT (B2), and co-operate with other organizations (B8). The EA-Induced Capabilities should be regarded as representing crucial intermediate benefits, mediating the effect between other determinants (e.g. Architectural Insight) and end goals. The end goal construct *Organizational Performance* represents the degree to which the organization is able to obtain EA-related end goals at the level of the organization as a whole. This comprises often-mentioned EA-benefits, viz. achieving enterprise-wide goals rather than goals of local units (B1), controlling costs (B6) and achieving organizational agility (B7). The end goal *Project Performance* is the degree to which the organization's individual projects are able to obtain benefits from EA. This comprises benefits often mentioned in the EA-literature, namely regarding budget (B12), deadlines (B13), quality (B15), functionality (B16), risk management (B14), project complexity control (B17) and speed of initialization (B18).

The construct *EA Approach* conceptually represents a part-of relationship because the techniques are separate components of the overall approach. The benefit constructs *Architectural Insight*, *EA-Induced Capabilities*, *Organizational Performance* and *Project Performance* all represent a type-of relationship since individual benefits are specific instances of these more general and abstract benefit constructs.

Figure 2 presents the final model. We discuss its statistical properties in section 4.3.2 and a detailed discussion in section 4.3.3.

## 4. RESEARCH DESIGN, DATA ANALYSIS AND RESULTS

In this section we present the research design, descriptive statistics and sample representativeness, followed by a discussion of the explanatory PLS model.

### 4.1. RESEARCH DESIGN

The development of our framework into a more mature research model was not only informed by existing literature and our research discussions, but was also constrained and bounded to reality by data from the 'real world'. This empirical component entails an online survey and subsequent statistical analysis of its perceptual measures, resulting in a partial least squares (PLS) model. The unit of observation is the individual worker's perspective on EA benefits and practices (Weinstein 2010). This allows for obtaining a balanced insight from multiple roles (enterprise architects, managers, business and system analysts, software architects, et cetera) instead of one, possibly biased, role (e.g. only managers; Venkatraman & Ramanujam 1987). The target population is defined as "all people working in The Netherlands in commercial or public organizations (either as internal employees or external consultants) who in a professional capacity have to deal with Enterprise Architecture (as an EA creator, an EA user or both)".

The design of the survey underwent several iterations. The questionnaire was first created as a written document by the principal researcher, which was reviewed by the other authors and a questionnaire expert from a statistical research institute. This led to improvements in the design, such as using question-dependent response categories and textual clarifications. The resulting questionnaire was used to create the web-survey, which was similarly reviewed. Finally, we verified the web-survey with three test-respondents (who worked for an IT service provider and a government agency). They had to fill in the questionnaire whilst articulating their thoughts out loud, so as to let us gain an insight into the minds of respondents. This yielded several minor simplifications and clarifications. The final questionnaire items (translated to English) can be found in Appendix A. See section 5 for a discussion on objective versus subjective measures.

Since we did not specifically target executive management, promising incentives such as benchmarking reports would not have increased the response rate or quality. In combination with the fact that long surveys do increase the probability of boredom and fatigue (Lindell & Whitney 2001), we opted for a relatively short questionnaire of 45 questions. Because of the above, the exploratory nature of our study, the absence of predefined EA-related measurement items for survey research, and the fact that we needed to incorporate a large number of distinct variables, it was not feasible to measure each individual practice and benefit reflectively with multiple items (cf. Drolet & Morrison 2001a, 2001b; see also sections 4.3.1 and 4.3.2 for the benefits of formative measurement).

All survey questions explicitly referred to the current (or latest) organization in which the respondents actually carried out their work, for example because they were an employee or because they were posted there as an external consultant. The term "Enterprise Architecture" was defined at the beginning of the survey. The first survey question explicitly asked whether the respondent has to deal with EA in his or her work (the survey was terminated if this was not the case). In general, questions featured five closed response categories (e.g. from *1. Very poor* to *5. Very good*) and one *No answer* category.

Since registers containing contact information of the individuals comprising our target population were not available, we used several communities related to information systems and architecture. First, an email containing the hyperlink to the web-survey was sent to relations and employees of several IT-service providers and IT-intensive organizations. Second, the web-survey was advertised at two IS architecture conferences in The Netherlands attended mainly by practitioners. Consequently, the survey requests have been sent to 2500 potential respondents. The data were gathered between October 2009 and May 2010.

In total, we received 293 valid surveys. A questionnaire had to pass three types of checks to be accepted as valid. First, it had to be completed and submitted. Second, it was verified whether there was a basic consistency between key variables. If the respondent's score on B11 was (very) poor, then B1, B2 and B7 were not allowed to all be scored (very) good. Likewise, if the respondent's score on B11 was (very) good, then B1, B2 and B7 could not be (very) poor. Third, since we did not work with website passwords, we performed basic duplicate records checks. However, only unique records were found, and it required merely 11 included variables for a query to return zero duplicates.

## 4.2. DESCRIPTIVES AND REPRESENTATIVENESS

The 293 respondents worked for 116 different organizations. Whereas general questions yielded high responses, several individual questions resulted in quite some item-nonresponse, especially regarding projects. The free-text option at the end of the survey provides clues as to why. For example, some Enterprise Architectures focused not on internal project decisions, but on high-level portfolio management instead. Furthermore, working with EA was relatively new in several organizations, making its evaluation difficult. Males dominate the field, 268 in total (91.5%), versus only 21 (7.2%) of the respondents being female and 4 (1.4%) unspecified. The tables below present the distribution of organizational roles. Due to potential differences in perceptions (see the ICIS paper mentioned in the Remarks for more on this), we considered it desirable to have a sample distribution that features a roughly equal number of EA users and EA creators. We conclude from Table 2 that this condition is satisfied.

| Table 1. Roles occupied by respondents (multiple roles allowed) | | | |
|---|---|---|---|
| **Role** | | **Frequency** | **Percentage** |
| EA Creator | Enterpr Architect Business & Information | 97 | 33.1% |
| | Enterpr Architect Application & Infrastructure | 95 | 32.4% |
| | Manager | 39 | 13.3% |
| | External EA Consultant | 19 | 6.5% |
| EA User | Manager | 42 | 14.3% |
| | Project Manager | 39 | 13.3% |
| | Project Architect | 56 | 19.1% |
| | Business Analyst/Designer | 34 | 11.6% |
| | System & Information Analyst/Functional Designer | 26 | 8.9% |
| | Software Architect | 35 | 11.9% |
| | Technical Designer | 19 | 6.5% |
| | Developer/Programmer | 8 | 2.7% |
| | Maintenance Engineer | 8 | 2.7% |

| Table 2. EA creators and users | | |
|---|---|---|
| **Role** | **Frequency** | **Percentage** |
| EA Creator | 107 | 36.5% |
| EA User | 109 | 37.2% |
| EA Creator and User | 65 | 22.2% |
| Unknown | 12 | 4.1% |
| Total | 293 | 100% |

| Table 3. Organizational size | | |
|---|---|---|
| **#Employees** | **Frequency** | **Percentage** |
| <2000 | 81 | 27.7% |
| 2000-5000 | 78 | 26.6% |
| ≥5000 | 128 | 43.7% |
| Unknown | 6 | 2.0% |
| Total | 293 | 100% |

We investigated the economic sectors to assess the representativeness of our sample. Since we could not use a pre-defined sampling frame corresponding with our target population (e.g. a population register of EA stakeholders), it was difficult to determine the extent to which our sample constitutes a representative subset. However, we could use previous research on EA for a comparison of sector distributions. Table 4 presents the distributions of our survey and those of two other studies. The column on the right (in white) presents the distribution found by Obitz & Babu (2009) in their survey with 173 respondents drawn from the global IT community (mainly North America). The middle column (in gray) contains our results. The left column presents the distribution found by Bucher et al. (2006).

| Table 4. Distribution of respondents across economic sectors | | | | | | |
|---|---|---|---|---|---|---|
| **Bucher et al. (2006)** | | | **Research in this paper** | | | **Obitz and Babu (2009)** |
| **Industry** | **%** | **%** | **Industry (based on ISIC Rev. 4)** | **%** | **%** | **Industry** |
| Finance and insurance | 62.5 | 61.8 | Financial and insurance activities | 30.4 | 30.4 | 28.4 | Banks, insurance, financial services and capital markets |
| Industry | 12.5 | 13.2 | Manufacturing (products and food) and construction | 5.5 | 6.5 | 4.9 | Industrial goods/engineering, food, beverage, tobacco, pharmaceuticals & biotech |
| | | | Agriculture, fishing, forestry and mining | 1.0 | | | |
| Software, IT and Telecommunication | 25 | 25.0 | Information, communication, entertainment and recreation | 12.3 | 12.3 | 16.1 | Media, information, entertainment, telecom services, travel and leisure |
| Total | 100% | 100% | Public administration (including defense) | 31.1 | 32.8 | 23.5 | Government, education, aerospace and defense |
| | | | Education and research | 1.7 | | | |
| | | | Energy & water supply and waste management | 5.1 | 5.1 | 6.2 | Oil, gas and utilities |
| | | | Human health and social work activities | 2.7 | 2.7 | 7.3 | Healthcare services |
| | | | Trade, transportation, hotel, catering, real estate and other services | 10.2 | 10.2 | 13.6 | Retail, transportation, logistics and other |
| | | | Unknown | 0 | 0 | N.a. | Professional services |
| | | | Total | 100% | 100% | 100% | Total |

We asked consultants posted at clients' offices to fill in the questionnaire from the perspective of their customer organization. We therefore have also proportionally dispersed the *Professional services* category of Obitz & Babu (2009) in order to obtain a better comparison. The left and right columns demonstrate that the various industry distributions of these two studies are fairly similar. Our collapsed *Public administration, education and research* category is large, however, compared to Obitz & Babu. This is consistent with the fact that the public administration sector in The Netherlands is large compared to that of North America. All in all, we interpret these similarities in distributions as support for having a representative sample.

### 4.3. AN EXPLANATORY PLS MODEL FOR EA PRACTICES AND BENEFITS

This section presents the PLS analysis of the model presented in Figure 2. We start by discussing several methodological considerations.

#### 4.3.1 METHODOLOGICAL CONSIDERATIONS

To get acquainted with the data, we started with simply measuring ordinal association using Kendall's tau-b, Spearman's rho and univariate and multiple ordinal regression analyses (cf. Norušis 2008, 2009). This yielded the first results, which took the form of an ordinal regression model at the level of individual practices and benefits (very much akin to Figure 1, see the ICIS paper). Subsequently, we have employed PLS to analyze the deeper causal mechanisms with a path model and to include aggregated (conceptually more abstract) constructs in order to focus on the theoretical essence.

We have used formative constructs to be able to investigate the substantive relationships at the desired level of abstraction. Using this type measurement allows for summarizing variables to gain rich theoretical concepts and nomological parsimony (Bagozzi 2011; Carver 1989; Cenfetelli & Bassellier 2009; Petter et al. 2007). This enables researchers to investigate more complex theoretical effects without creating a chaotic web of relationships at too detailed an abstraction level. The following considerations show that all constructs in our study should be regarded as formative instead of reflective (cf. Jarvis et al. 2003; Petter et al. 2007; Roberts & Thatcher 2009). All of the indicators measure conceptually distinct aspects of the higher-level construct. As such, they are not interchangeable and they do not necessarily have to covary with each other. The indicators define the construct – as opposed to reflect it – and changes in the indicators are expected to 'determine' the changes in the construct (not vice versa). We view a construct as defined by – and thus summarizing – the distinct indicators. In other words, the indicators are part of the construct's definition and the construct is a definitional abstraction of its indicators. Our research therefore requires purely 'vertical aggregation', which means that a construct's weights are only determined by its own indicators. In the alternative 'good neighbor' approach (cf. Adelman & Lohmöller 1994; Kock 2011) a construct can get contaminated because its score is dependent on items of other constructs in the nomological model. This is undesirable, as the construct is no longer a pure abstraction of its underlying aspects or indicators. In addition, this contamination can result in interpretational confounding, unstable weights, higher levels of collinearity, inflated path coefficients and point variable instability (Bagozzi 2011; Bollen 2007; Cenfetelli & Bassellier 2009; Esposito Vinzi et al. 2010; Howell et al. 2007; Kim et al. 2010; Kock 2011).

We have used WarpPLS 3.0 (Kock 2012) for several reasons. First, this multivariate analysis tool allows for identifying nonlinear relationships. Second, WarpPLS is particularly well-suited for models with formative constructs because they are calculated without other constructs contaminating their weights (by using the PLS regression algorithm). The non-linear Warp3 algorithm was employed for the inner model, allowing for sigmoidal shapes (S curves). Bootstrapping was used with 250 resamples.

### 4.3.2 TECHNICAL EVALUATION OF THE MODEL

This section presents the evaluation of the inner and outer model. This evaluation is discussed in detail because of the exploratory nature of our study. In terms of *indicator-level validity*, Table 5 shows that each indicator weight is statistically significant (with almost all p-values being <0.001 with both jackknifing and bootstrapping). This means that all indicators have a significant positive relative contribution to their respective construct (Cenfetelli & Bassellier 2009). The same holds for the loadings, which represent the absolute importance of the indicator to the formative construct (ibid.; Hair et al., 2011). The VIF (Variance Inflation Factor) for construct indicators shows how redundant its information is (Urbach & Ahlemann 2010, Kock 2012) and should be lower than 2.5 (Kock 2012). This study's indicator-level VIF's all have values below 2.1 (see Table 5), meaning that multicollinearity among the constructs' indicators is not an issue.

In terms of *construct-level validity*, we can first note that composite reliability and Cronbach's alpha are not relevant criteria in a formative context. However, item B11 is a representation of EA benefits in general (see Appendix A) and should thus be strongly correlated to Organizational Performance, Architectural Insight, EA-Induced Capabilities and Project Performance. With each p<0.001 and coefficients of 0.700, 0.655, 0.671 and 0.375 respectively, this is indeed the case. Second, the full collinearity VIF for vertical and lateral collinearity should be below 3.3 for each construct (Kock & Lynn 2012). The highest value being 2.533, all of our constructs pass this test. Third, vertical collinearity was measured by calculating block VIFs for each construct with two or more predictors. With a maximum block VIF of 1.934 in our model, all values are clearly below the 3.3 threshold (Kock 2012). Another important criterion for formative models is nomological validity, which means that the formative constructs should be included as part of a model and do not behave unexpectedly (Roberts & Thatcher 2009; Urbach & Ahlemann 2010). We argue that this is indeed the case (see sections 3.3 and 4.3.3). Lastly, a requirement for formative constructs is that they cover all underlying facets and thus form the concept space (Andreev et al. 2009; Kim et al. 2010; Petter et al. 2007). We have dealt with this content coverage demand by identifying the EA practices and benefits in the EA literature (see section 3.2 to 3.4). The authors, coming from academia and practice, have also continuously discussed the meaning and validity of the constructs (Hair et al. 2011).

| | Table 5. Construct weights, standard errors, P-values (bootstrapping) and indicator VIF's | | | | | | | | |
|---|---|---|---|---|---|---|---|---|---|
| Item | | EA Approach | Project Compliance | Architectural Insight | EA-Induced Capabilities | Organizational Performance | Project Performance | SE | P-value | VIF |
| T1 | Formal approval of EA | (0.131) | 0.000 | 0.000 | 0.000 | 0.000 | 0.000 | 0.020 | <0.001 | 1.168 |
| T2 | EA choices linked to organization | (0.166) | 0.000 | 0.000 | 0.000 | 0.000 | 0.000 | 0.021 | <0.001 | 1.376 |
| T3 | Management propagation of EA | (0.182) | 0.000 | 0.000 | 0.000 | 0.000 | 0.000 | 0.016 | <0.001 | 1.534 |
| T4 | Compliance assessments | (0.205) | 0.000 | 0.000 | 0.000 | 0.000 | 0.000 | 0.016 | <0.001 | 1.788 |
| T5 | Knowledge exchanges architects | (0.188) | 0.000 | 0.000 | 0.000 | 0.000 | 0.000 | 0.020 | <0.001 | 1.694 |
| T6 | Knowledge exchanges projs/architects | (0.170) | 0.000 | 0.000 | 0.000 | 0.000 | 0.000 | 0.020 | <0.001 | 1.562 |
| T7 | Providing assistance | (0.185) | 0.000 | 0.000 | 0.000 | 0.000 | 0.000 | 0.018 | <0.001 | 1.447 |
| T8 | PSA | (0.151) | 0.000 | 0.000 | 0.000 | 0.000 | 0.000 | 0.020 | <0.001 | 1.379 |
| T9 | Document templates | (0.176) | 0.000 | 0.000 | 0.000 | 0.000 | 0.000 | 0.018 | <0.001 | 1.528 |
| T10 | Finance (dis)incentives | (0.076) | 0.000 | 0.000 | 0.000 | 0.000 | 0.000 | 0.023 | <0.001 | 1.098 |
| O1 | Project Compliance | 0.000 | (1.000) | 0.000 | 0.000 | 0.000 | 0.000 | 0.000 | <0.001 | 0.000 |
| B3 | Understanding organizat. complexity | 0.000 | 0.000 | (0.383) | 0.000 | 0.000 | 0.000 | 0.033 | <0.001 | 1.454 |
| B9 | Clear image of future | 0.000 | 0.000 | (0.405) | 0.000 | 0.000 | 0.000 | 0.030 | <0.001 | 1.596 |
| B10 | Communicating effectively | 0.000 | 0.000 | (0.376) | 0.000 | 0.000 | 0.000 | 0.038 | <0.001 | 1.384 |
| O2 | Shared interpretation EA norms | 0.000 | 0.000 | (0.195) | 0.000 | 0.000 | 0.000 | 0.054 | <0.001 | 1.050 |
| B2 | B/IT alignment | 0.000 | 0.000 | 0.000 | (0.344) | 0.000 | 0.000 | 0.027 | <0.001 | 1.661 |
| B4 | Controlling organizational complexity | 0.000 | 0.000 | 0.000 | (0.314) | 0.000 | 0.000 | 0.031 | <0.001 | 1.409 |
| B5 | Integration, standardization, deduplic. | 0.000 | 0.000 | 0.000 | (0.329) | 0.000 | 0.000 | 0.028 | <0.001 | 1.544 |
| B8 | External co-operation | 0.000 | 0.000 | 0.000 | (0.315) | 0.000 | 0.000 | 0.030 | <0.001 | 1.408 |
| B1 | Enterprise-wide goals | 0.000 | 0.000 | 0.000 | 0.000 | (0.448) | 0.000 | 0.034 | <0.001 | 1.510 |
| B6 | Controlling costs | 0.000 | 0.000 | 0.000 | 0.000 | (0.381) | 0.000 | 0.043 | <0.001 | 1.215 |
| B7 | Agility | 0.000 | 0.000 | 0.000 | 0.000 | (0.439) | 0.000 | 0.031 | <0.001 | 1.466 |
| B12 | Project budgets | 0.000 | 0.000 | 0.000 | 0.000 | 0.000 | (0.229) | 0.032 | <0.001 | 2.083 |
| B13 | Project deadlines | 0.000 | 0.000 | 0.000 | 0.000 | 0.000 | (0.221) | 0.031 | <0.001 | 2.056 |
| B14 | Project risk management | 0.000 | 0.000 | 0.000 | 0.000 | 0.000 | (0.205) | 0.032 | <0.001 | 1.486 |
| B15 | Project quality | 0.000 | 0.000 | 0.000 | 0.000 | 0.000 | (0.252) | 0.030 | <0.001 | 1.960 |
| B16 | Project functionality | 0.000 | 0.000 | 0.000 | 0.000 | 0.000 | (0.237) | 0.034 | <0.001 | 1.719 |
| B17 | Project complexity control | 0.000 | 0.000 | 0.000 | 0.000 | 0.000 | (0.239) | 0.028 | <0.001 | 1.699 |
| B18 | Speed of project initialization | 0.000 | 0.000 | 0.000 | 0.000 | 0.000 | (0.073) | 0.033 | 0.015 | 1.049 |

Note that this table provides a concise overview of which items belong to which construct.

Regarding *model-level validity*, three fit indices were calculated. The APC (Average Path Coefficient), ARS (Average $R^2$) and AVIF (Average Variance Inflation Factor) are shown in Figure 2. The APC and ARS are both highly significant, with P-values of <0.001. The AVIF, with a value of 1.438, is lower than the suggested maximum value of 5 (Kock 2012). Furthermore, all $Q^2$ values (presented in Table 6) are well above 0, indicating good predictive relevance (Chin 2010; Kock 2012). The $f^2$ effect sizes (see Table 7) range from small to large (Cohen 1988).

| Table 6. Full collinearity VIF's and $Q^2$ values | | | | | | |
|---|---|---|---|---|---|---|
| Measure | EA Approach | Project Compliance | Architectural Insight | EA-Induced Capabilities | Organizational Performance | Project Performance |
| Full collinearity VIF | 1.432 | 1.294 | 2.131 | 2.521 | 2.533 | 1.191 |
| $Q^2$ value | N.A. | 0.214 | 0.236 | 0.476 | 0.599 | 0.143 |

| Table 7. Cohen's $f^2$ effect sizes | | | | | |
|---|---|---|---|---|---|
| Construct | EA Approach | Project Compliance | Architectural Insight | EA-Induced Capabilities | Project Performance |
| Project Compliance | 0.214 | | | | |
| Architectural Insight | 0.148 | 0.083 | | | |
| EA-Induced Capabilities | | 0.057 | 0.420 | | |
| Organizational Performance | | | 0.182 | 0.359 | 0.057 |
| Project Performance | | | 0.069 | 0.071 | |

WarpPLS indicated that 8 out of 10 structural relationships were non-linear. However, none of the relationships showed a statistically significant difference between the non-linear coefficients and their linear counterparts (using the two methods described in Kock 2014). We therefore conclude that these relationships were only slightly curved (see Appendix A for an example). It was also verified whether interaction effects exist, which could increase the explanatory power of the model. However, since the moderators found were only barely significant and did not invite a meaningful theoretical interpretation, they were not included in the final model. Finally, all structural relations were checked for confounding effects by controlling for the influence of five control variables: *organizational size, economic sector, EA focus, number of software and project architects* and *number of domain and enterprise architects*. However, all relationships retained their significant effects, regardless of whether or not these contextual variables themselves significantly contributed to the model.

Formative indicators are relatively independent (Jarvis et al. 2003; Petter et al. 2007) and not necessarily required to have conceptual unity (Bollen 2011). This allows – and arguably even necessitates (Carver 1989) – extending the analysis of *relationships between formative constructs* (i.e. between aggregated variables) with the analysis of *underlying individual relationships* (i.e. between indicators of separate constructs). This is certainly relevant in our study, since we regard formative constructs as an abstraction of their indicators. Individual indicators thus retain their right of existence. Note that the individual relationships in our study cannot be investigated by analyzing the constructs' weights. In general, the inter-item causal effects tend to be significant here, meaning that each item is significantly related to one or more items of other constructs, even when taking a critical approach (see Appendix B). However, below we will discuss some cases of non-significant relationships.

We have also tested for common method bias. First, SPSS 17.0 was used to perform Harman's single factor test on the constructs' indicators, which revealed that several factors existed in our dataset and that one general, single factor could not account for the majority of the covariance. Second, the smallest correlation and, more conservatively, the second smallest positive correlation among the manifest variables can be seen as proxies for common method variance (Bagozzi 2011; Lindell & Whitney 2001). In our study these correlations were 0.001 and 0.002 respectively, and were thus not indicative of any problems. Finally, the full collinearity test already discussed was used as another common method bias test for variance-based SEM (Kock & Lynn 2012). As a final general validation, a second and wholly different PLS analysis was conducted with equal indicator weights (summed scales) and estimation of linear relationships. Although most path coefficients and $R^2$ values were slightly lower, however, all relationships remained highly statistically significant, thus providing additional support for the robustness of our model.

Our iterative approach included regularly discussing the results, analyzing survey data and frequently revisiting the literature. This led to the inclusion of various intermediate constructs. Architectural Insight was added first, due to many publications on EA emphasizing the importance of architectural modeling and insight. For a certain time the Organizational Performance construct also included the benefits that now constitute EA-Induced Capabilities. However, as complexity management and standardization cannot be considered end goals, we decided that such intermediate EA benefits should be acknowledged in the form of a separate construct. This resulted in the definition of EA-Induced Capabilities, which greatly increased the explanatory power of the model. Several alternative relationships for which theoretical justification could be provided were also analyzed, but were not included because no statistical justification could be found or because of parsimony reasons. In this context, we concluded that the EA Approach did not contribute directly to the end goals Project Performance (due to a statistically insignificant relationship) and Organizational Performance (due to the lack of a notable increase in $R^2$, viz. from 0.598 to 0.610). Direct effects from Project Compliance to these two end goals were both insignificant and therefore not included in the model. Finally, a significant direct relationship between EA Approach and EA-Induced Capabilities was found, but also not included in the final model due to the modest increase in $R^2$ it brought with it (viz. from 0.477 to 0.500). This also implies our model suggests *full mediation* for predictors EA Approach and Project Compliance with regard to dependent variables Project Performance and Organizational Performance (Shrout & Bolger 2002). This is mainly due to insignificant direct paths when controlling for the indirect paths. As noted, however, for some paths a case for partial mediation could in principle be made (also see our suggestions for future research). The theoretical discussion on mediation can be found in section 4.3.3.

### 4.3.3 THEORETICAL DISCUSSION OF PLS RESULTS

This section presents the theoretical discussion of the model. Note that the focus is on the model itself and that consistency with related literature is discussed in sections 3.2 and 3.3. It can be seen from Figure 2 that all significant relationships at the construct level have positive signs. In general we can thus conclude that using the EA practices of the EA approach directly results in more compliance and architectural insight, which in turn lead to increased organizational capabilities and more benefits at both the project and the organizational level. We discuss these results in more detail below. The individual item level is referred to where relevant, in order to present the reader with tangible interpretations and salient details (based on the tables of Appendix B).

The *EA Approach* exerts an indirect influence on the organization's end goals, but has a direct positive impact on the intermediate results Project Compliance with EA and Architectural Insight, which play a central role in gaining benefits. In this context, knowledge exchanges between architects and project members result in an increased ability to work in a compliant fashion. Furthermore, having management emphasize the importance of EA and announcing compliance assessments increases a project's motivation and commitment to become compliant, possibly to avoid confrontation. The EA Approach also directly and significantly results in improved Architectural Insight, regardless of whether a project complies or not. This holds for practices such as knowledge exchanges between architects (which directly aim at knowledge sharing) and ensuring the EA is explicitly linked to the strategic business goals (which makes clear what the most important objectives are and how they are related to the internal organization). See section 4.3.4 for a more detailed analysis of the effect of individual EA practices.

*Project Compliance with EA* is an important intermediate variable and mediates the effect of EA practices. Compliance is not intrinsically beneficial, as its effect depends on the quality of the EA. However, since EAs typically feature insight into the organization's systems and processes (which improves decision making) and include proven generic best practices (e.g. modularity, service orientation, loose coupling, patterns), compliance can be expected to generally yield positive effects. Compliance works as a mediator and no empirical indications of moderation were found. The EA Approach has a positive effect on compliance, which in its turn exerts a significant positive influence on Architectural Insight and EA-Induced Capabilities. The reason for the first effect is that compliance implies that EA prescriptions are used not only actively, but also correctly. This results in increased understanding because the project members get familiarized with the EA content (including holistic insights and knowledge) and they obtain relevant practical experiences as a result. There is also an effect of Project Compliance on EA-Induced Capabilities. This represents leveraging the value of routine modes of operation and standardized technologies, functionality and data definitions. This enables not only standardization, integration and deduplication of processes and systems, but also easy data sharing and different forms of co-operation with other organizations. In addition, it makes it easier to manage complexity.

In its turn, *Architectural Insight* has an effect on EA-Induced Capabilities, Project Performance and Organizational Performance. These three relationships essentially represent informed and effective decision making regarding which processes and systems to implement, integrate, standardize or deduplicate (by management), as well as how to pursue these initiatives (by projects). A shared vision results, for example, in consistent behavior and alignment between different levels. Knowledge of the organization also yields optimized resource and project portfolio decisions, such as which projects to initiate or to cancel. Furthermore, architectural insight into complexity is a prerequisite for actually managing this complexity. The views provided by EA, covering all relevant aspects and providing insights into various aspects, also lead to more realistic decisions and design choices that are coherent both within and between projects. Moreover, the views should result in projects having to deal with fewer surprises and being better capable of dealing with risk, complexity and scope creep. Finally, stakeholders can communicate more easily, as a result of the EA's definitions and frames of reference. This should have a positive impact on preventing conflicts, encouraging cooperation, and resolving e.g. complexity and redundancy. Like compliance, Architectural Insight works as a mediator and has no moderating function.

*EA-Induced Capabilities* directly affect Project Performance and Organizational Performance. The EA-Induced Capabilities construct represents the outcomes of the organization's EA that, although beneficial in themselves, have a foundational role in obtaining EA-related end goals. Improved capabilities imply increased expertise and know-how of employees, as well as effective organizational patterns and modes of working. Being able to align business and IT integrate processes and systems enables the organization to increase productivity and quality, while standardization and deduplication results in more efficiency. The capability to control complexity and increase standardization also mitigates risks, resulting in an increased ability to achieve both project and organizational goals.

Another capability is co-operating with other organizations, e.g. in the form of co-sourcing or strategic partnerships. Because of the resulting acquired expertise, innovation and improved modes of working, this can yield further organizational benefits such as agility, cost reductions, project success and new products and services.

*Project Performance* can be regarded both as an end goal (at least for local organizational units) and as an instrument for achieving the broader, enterprise-wide goals of Organizational Performance. This is the result of the fact that project outcomes are often a necessity for realizing organizational goals. For example, delivering innovative new products and services is usually dependent on their implementation by projects. Moreover, projects that improve existing processes can directly contribute to agility and lower costs.

*Organizational Performance* represents the end goals of EA at the level of the entire organization. These are the typical EA-related goals often mentioned in the literature, such as efficiency and agility. Organizational Performance is affected by Architectural Insight (as a result of improved decision making), EA-Induced Capabilities (due to improved processes, systems and co-operation) and Project Performance (because of projects that contribute directly to the enterprise's end goals).

Interestingly, projects in general do not benefit as much from working with EA as the entire organization does. This is indicated by the striking difference between effects on Organizational Performance and Project Performance in terms of $R^2$s (0.598 and 0.140 respectively). The smaller effect magnitudes are another indication, as demonstrated by the beta coefficients in Figure 2 and the effect sizes in Table 7. Note that adding the constructs EA Approach and Project Compliance as predictors of Project Performance neither yield significant relationships nor a notable increase in the explained variance of this construct. The reason for projects benefiting less may be that EA aims primarily at achieving enterprise-wide goals (Richardson et al. 1990; Lankhorst et al. 2005), and, despite some grand claims, ultimately focuses less on the results of individual projects. In any case, these empirical findings clearly show that a distinction between organizational and project performance is highly relevant in a theoretical model of EA practices and benefits.

Although all relationships remained statistically significant after controlling for the contextual variables, the economic sector proved to have significant effects itself when included. Generally, the *governmental* and *financial* sectors perform slightly poorer than the other sectors in terms of Project Compliance, Architectural Insight and EA-Induced Capabilities. Governmental organizations also perform slightly worse in terms of Organizational Performance. One explanation for the consistent underperformance of governmental organizations may be that this sector invests significantly less in its EA Approach than the other sectors (this is indeed the case in our sample), which according to our causal model should manifest itself in lower scores of the benefit constructs.

### 4.3.4 THE MOST EFFECTIVE EA PRACTICES

Since the organization can directly invest in the practices that constitute the EA Approach, it is interesting to investigate them here in more detail and in relationship with each other. We have therefore analyzed which practices have statistically significant relationships with Project Compliance and Architectural Insight. We are allowed (or even required, cf. Carver 1989; Edwards 2011) to conduct such an analysis because these indicators of EA Approach represent different concepts and we use formative constructs here mainly to achieve a higher level of theoretical abstraction (which does not deny the indicators their right of individual existence).

The results are presented in Figure 3, which essentially zooms in on the left part of the model in Figure 2. Project Compliance with EA is significantly affected by several practices. All relationships but one are positive, thus the more a technique is used, the higher the achieved level of conformance will generally be. Being assessed on compliance (T4) has the strongest effect on whether projects will actually comply. The fact that a project will be explicitly confronted with its nonconformance apparently stimulates its members to conform to the architectural norms. This could be due to their desire to avoid confrontation, or to the fact that carrying out assessments is an indication of the importance of compliance. This latter mechanism will surely play a role in management propagation and stimulation of EA (T3), the practice with the second strongest influence. Knowledge exchanges between architects and project members (T6) also have a significant influence on Project Compliance with EA. This can be explained by the fact that a necessary condition for becoming compliant is to understand the norms.

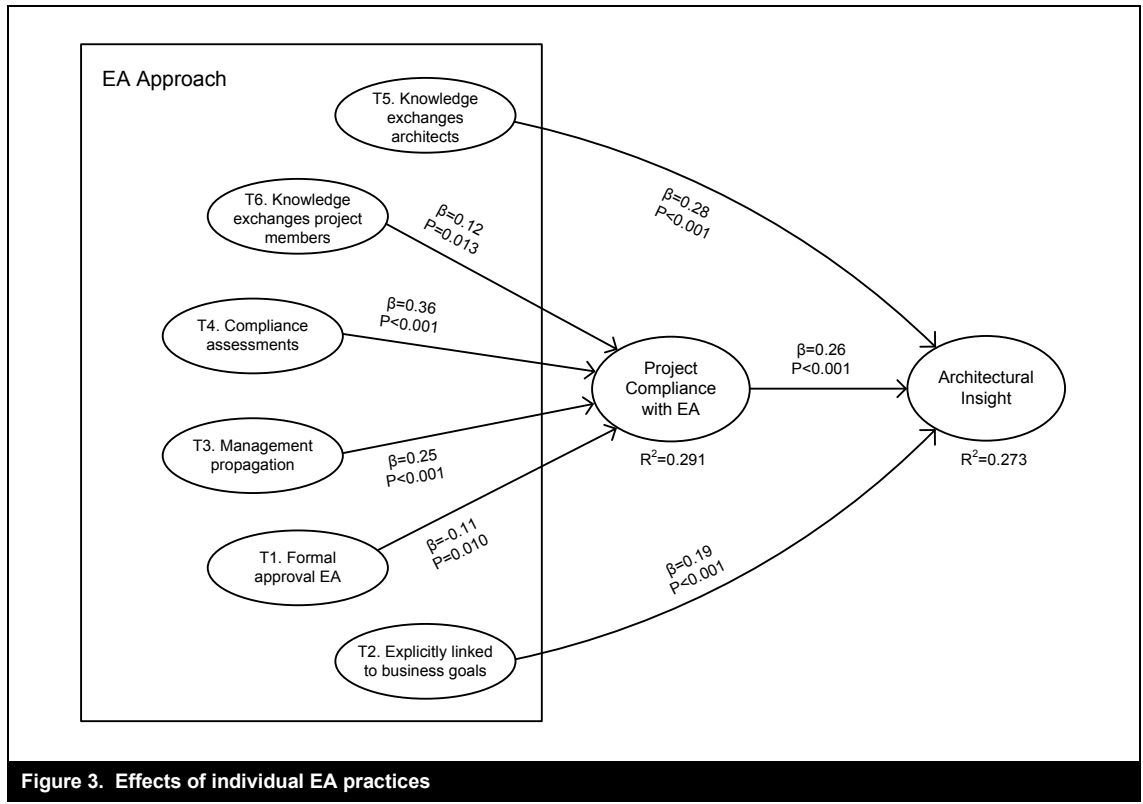

**Figure 3. Effects of individual EA practices**

An unexpected finding is that formally approving the EA (T1) has a statistically significant negative impact on compliance when it is controlled for other factors. A plausible explanation is that formal approval brings with it many time-consuming *bureaucratic procedures*, which can be expected to have a negative effect on achieving EA compliance and benefits, and possibly also lead to a less pragmatic attitude towards problem solving. An alternative explanation may be *complacency*, since organizational actors may assume all will turn out satisfactorily once matters are formalized. This can have a paralyzing effect that may very well be amplified by the perceived loss of freedom. Providing assistance to projects applying the EA's norms (T7) seems to have an ambiguous effect. It does not have a significant relationship in this model, but does have so in other models.

Architectural Insight is affected by three factors, one of which is Project Compliance. The two EA practices with a clear-cut effect on insight are organized knowledge exchanges between architects (T5) and explicitly linking the decisions in the EA to the business goals of the entire enterprise (T2). The effect of the former is self-explanatory, since such exchanges will spread, increase and enrich the knowledge about the EA and the organization to which it pertains. There is also a good reason why T2 has a significant effect on Architectural Insight: linking the choices in the EA explicitly to the strategic business objectives communicates a clear signal about the organization's priorities. Practices that focus on project artifacts (T8 and T9) do not seem to have a notable effect on the two endogenous constructs.

Assessing projects on compliance with EA (T4) and knowledge exchanges between architects and project members (T6) have a direct significant impact on Project Compliance. However, they only have a direct significant relationship with Architectural Insight when compliance is not included in the model (see Table C.1 of Appendix B for the precise relationships). When Project Compliance is added and controlled for in the model of Figure 3, both relationships are rendered insignificant (with p-values of 0.101 and 0.064 respectively). This means that the effects of these two practices are *fully mediated* by compliance (Shrout & Bolger 2002). This makes sense, since the mediating variable Project Compliance – like the practices affecting it – operates at the level of projects. Consistent with this observation is the fact that knowledge exchanges between architects (T5), operating at the level of the organization rather than that of projects, have a significant effect on Architectural Insight instead of on Project Compliance. More in general, all structural relations were checked for confounding effects by controlling for the five contextual variables, but all relationships retained their statistical significance regardless of them. Finally, if we merely include the six effective practices of Figure 3 in the EA Approach construct of Figure 2, the coefficients of the latter model change only slightly.

Although we have not seen a model similar to Figure 3 in the literature, our results are consistent with related publications. Schmidt and Buxmann (2011) – although not studying the causal chain of *how* EA delivers value – also find that a formal review and approval process for change projects is the most important EA success factor. Moreover, they conclude that communication and support, which are strongly related to our knowledge exchanges, are important determinants of success. Finally, they too find only weak support for the value of EA artifacts.

## 5. CONCLUSIONS AND FURTHER RESEARCH

Given the paucity of explanatory and quantitative research in the field of EA, this exploratory study has yielded several contributions by iteratively employing existing theory, multivariate analysis of empirical data, and reflective discussions. Using formative constructs to focus on fundamental theoretical effects at an overall level, it is made explicit how the EA approach, project compliance, architectural insight and EA-induced capabilities play important roles in obtaining value from EA for both projects and the entire organization. The resulting model provides clear insight into how EA delivers value, and not only is consistent with existing theory but also synthesizes fragmented lines of thinking. Our explanatory model shows that EA and EA practices do not achieve end goals directly but work through intermediate outcomes, namely project compliance with EA, architectural insight and EA-induced capabilities. Project compliance represents the active and correct use of EA norms in change initiatives. This should be taken explicitly into account in causal models, since simply having an EA does not directly yield organizational results. Architectural insight represents the knowledge and understanding that EA offers to both management (for high-level decision making) and projects (for designing systems and processes). EA-induced capabilities refer to the abilities to implement foundational EA-related benefits, such as controlling complexity and integrating processes. We are not aware of any studies that include all these constructs in a single theoretical or statistical model. Projects and compliance, while having central mediating roles and contributing to organizational performance, are usually ignored in EA research. Practitioners too should give this issue sufficient attention. Projects benefit less from EA than the organization as a whole, but their success is often crucial for local organizational units.

Another contribution is the identification of efficacious EA practices. Although these techniques are known from the literature, their importance regarding project compliance with EA and architectural insight has to the best of our knowledge not yet been empirically investigated. Our study has identified four individual practices as important determinants of compliance, with compliance assessments and management propagation clearly having the largest effects. Knowledge exchanges between architects and project members also have a positive effect on compliance, whereas the formal approval of EA has a negative influence. Furthermore, architectural insight is affected by compliance and two practices, namely explicit linking of EA with business goals and knowledge exchanges between architects. Finally, we showed that the economic sector also has a small influence on EA performance, with governmental and, to a lesser extent, financial organizations benefitting less from architecture.

Owing to the recent methodological debate on formative measurement, we have circumvented potential drawbacks of formative constructs, such as interpretational confounding and point variable instability. This allowed us to demonstrate a very useful application of formative measurement – namely to focus on the general and theoretically substantive explanatory relationships at an optimum level of abstraction – while still being able to drill down to the details. More in general, we have shown how an exploratory or question-driven statistical analysis can yield valuable theoretical contributions.

In terms of implications, it is now clear that causal mechanisms in the field of EA are not simple. Causes and direct and indirect effects (benefits) should therefore be modeled explicitly. In such research endeavors it is important to determine the optimal degree of theoretical and methodological abstraction, in order to avoid both oversimplification and detail overload. We have identified and empirically supported several important EA practices and benefits. Future EA studies should explicitly model and test these constructs, including the usually ignored role of projects and compliance.

## Limitations and future research

Although our dataset and model demonstrate meaningful, consistent and statistically valid results, there are several limitations to consider. First, we have measured perceptions of individual respondents instead of objective facts. However, this is often the case with evaluative statistical research and perceptions have long ago been established as a valid indicator of organizational performance (Dess & Robinson 1984; Schryen 2013; Venkatraman & Ramanujam 1987; Wall et al. 2004). Furthermore, using so-called objective measures was not feasible in our study because of their fundamental shortcomings, such as definitions that differ between organizations in such a way as to threaten comparability (ibid.). For example, objective measurement would assume that all organizations in our sample have used the exact same definitions and measures in the previous years. Although this is indeed possible in some studies using legal definitions of profit and turnover (assuming individual accounting practices and fraud do not yield biased results), this was not an option for our study. One reason for this is the fact that we also study governmental organizations, which do not have profit. Furthermore, we are interested in EA-specific concepts, which are not covered in laws, regulations or other standardized approaches used in practice. For instance, objectively measuring 'degree of process integration' or 'risk' is not possible because each organization will have its own subjective measure for these concepts – assuming it is measured in the first place. Finally, even if all of the above would be of no concern, we would probably not have been able to find over a 100 organizations willing to spend considerable time and effort to provide this information. In short, we consider perceptions to be the optimal form of measurement for our study. However, future research could attempt to move from an analysis of individual perceptions to an analysis of projects, EAs and organizations by using objectively measured variables where possible as well as an explicit multilevel statistical model (cf. Hox 2002; Bélanger et al. 2014). Although multilevel analyses are rarely conducted in IS, they can improve the accuracy of the results (ibid.). Researchers should note, however, that this will result in a significantly more complex research project, especially when dealing with concepts that are difficult to uniquely identify in a survey (e.g., whereas organizations will be relatively easily identified, this will not be the case with individual projects).

A second limitation of our study is that the usual drawbacks of causal analysis based on observational rather than experimental data also apply to our study. It would require a randomized controlled experiment to prove causality in a more definitive manner. Unfortunately, this is not feasible when dealing with organizations. It is simply impossible to fully control for non-EA factors, such as organizational culture and economic crises. That being said, the ceteris paribus nature of variance-based SEM (partial coefficients) is a valuable approach for simulating laboratory settings (Wooldridge 2012). A third limitation is that our model needs to be tested in future research. Our exploratory study did not start with testable hypotheses, but generated them. Given the current state of EA theory this is appropriate, but confirmatory research with a different sample remains necessary. Do note that, for an exploratory analysis, this study does not run an unacceptable risk of capitalization on chance. Although the additional analysis of single-items in Figure 3 can be said to be somewhat vulnerable, the risk is limited in the core of our study, namely the model of Figure 2 (which uses aggregated constructs). Furthermore, the risk is limited due to the very low p-values observed throughout our study. Moreover, we did not employ automated model building methods (e.g. stepwise regression) – which are particularly vulnerable to overfitting – and have used bootstrapping procedures in our analysis (Babyak 2004). In this context it should also be noted that including many different variables in a study – as we did – effectively allows for statistically controlling for other factors (whereas a theory testing study with a 'sharp' focus brings with it the severe risk of not discovering confounders).

Despite the above limitations, we nonetheless consider this a valuable contribution to the field of EA. Not only because it offers important insights into the factors that determine the use and effectiveness of EA, but also because no other study seems to quantitatively model several important concepts at all. Note that our findings are also highly relevant for practitioners and for other forms of conformance, such as regulatory compliance. Finally, the reader should be aware of the fact that the identified positive and statistically significant relationships do not necessarily imply that the involved benefits are achieved to a "satisfactory" level (i.e. the respondents may not have opted for the two positive response levels often, e.g. *4. Good* and *5. Very good*). Although this is a generally neglected issue in explanatory research, we did investigate this important question and came up with nuanced results. We refer the interested reader to the ICIS paper that formed the basis for the current study (see Remarks).

We have three additional suggestions for future research. First, as the causal variables only explain part of the variance of the benefits, more factors should be taken into account when testing the model. For example, the type or size of projects. Time lags, which should be expected in EA, can also be accounted for explicitly. In addition, the ambition and nature of the EA itself can be incorporated, as very large top-down architectures seem to have a tendency to fail and introduce new significant risks (Ciborra & Osei-Joehene 2003; Grisot et al. 2014; Hanseth et al.

2006). Second, we did not include some significant paths for parsimony reasons, although some theoretical and empirical support could be brought forward (see section 4.3.2). These alternatives might thus be an avenue for future research. A third suggestion is to improve on the constructs. A more sophisticated concept of compliance could be used in this context to gain more fine-grained insights. For example, compliance could be measured by its four dimensions, namely correctness, justification, consistency and completeness (Foorthuis 2012). Furthermore, the techniques PSA (T8) and financial sanctions (T10) could be considered for removal from EA Approach, as they do not have any significant influence when controlled for other practices (see Table C.1 of Appendix B). The role of formal approval (T1) should also be reconsidered, since it consistently has a negative influence when controlled for other practices. Regardless of the specifics of future research, this study has clearly shown that EA offers different kinds of value, but that additional effort is required from the IS community to further investigate and confirm our findings.

**Remarks**. The preliminary results of this research project have been presented at ICIS 2010 as "On Course, But Not There Yet: Enterprise Architecture Conformance and Benefits in Systems Development".

**Acknowledgements**. The authors wish to thank Rik Bos, Jurriaan van Reijsen, Verena Dräbing, Nico Brand, Anne-Francoise Rutkowski, Joe Nandhakumar and the reviewers for their valuable remarks.

# APPENDIX A: QUESTIONNAIRE ITEMS AND EXAMPLES

## 1. Key questionnaire items

Regarding the architecture approach…

T1. The EA is formally approved (i.e. by line management).
T2. The choices made in the EA are explicitly linked to the business goals of the enterprise as a whole.
T3. Management propagates the importance of EA.
T4. Projects are being explicitly assessed on their degree of compliance with EA. [Note: this concerns the number of projects being judged on compliance (not the number of times one project is being assessed).]
T5. There is an organized knowledge exchange between different types of architects (for example enterprise, domain, project, software and infrastructure architects).
T6. There is an organized knowledge exchange between architects and other employees participating in projects that have to conform to EA (for example project managers, functional designers, developers and testers).
T7. Assistance is being offered in order to stimulate conformance to EA. (For example enterprise architects or change managers who help projects to make new designs conform to EA.)
T8. Projects make use of a PSA (Project Start Architecture).
T9. Document templates are being used to stimulate conformance to EA. (For example templates that focus attention on the EA by means of guiding texts and by requiring filling in relevant information.)
T10. Financial rewards and disincentives are being used in order to stimulate conformance to EA. (For example by covering the IT-expenses of a project if the solution is designed and built conform EA, or by imposing a fine for non-conformance.)

EA turns out to be a good instrument to…

B1. …accomplish enterprise-wide goals, instead of (possibly conflicting) local optimizations.
B2. …achieve an optimal fit between IT and the business processes it supports.
B3. …provide an insight into the complexity of the organization.
B4. …control the complexity of the organization.
B5. …integrate, standardize and/or deduplicate related processes and systems.
B6. …control costs.
B7. …enable the organization to respond to changes in the outside world in an agile fashion.
B8. …co-operate with other organizations effectively and efficiently.
B9. …depict a clear image of the desired future situation.
B10. EA turns out to be a good frame of reference to enable different stakeholders to communicate with each other effectively.
B11. EA, in general, turns out to be a good instrument.

Projects that have to conform to EA turn out to…

B12. …exceed their budgets less often than projects that do not have to conform to EA.
B13. …exceed their deadlines less often than projects that do not have to conform to EA.
B15. …deliver the desired quality more often than projects that do not have to conform to EA.
B16. …deliver the desired functionality more often than projects that do not have to conform to EA.
B14. …be better equipped to deal with risks than projects that do not have to conform to EA.
B17. …be better equipped to deal with complexity (of the project and/or its immediate environment) than projects that do not have to conform to EA.
B18. …get initialized faster than projects that do not have to conform to EA.

O1. Projects that are required to conform to EA turn out to actually conform to the architectural principles, models and other prescriptions.
O2. Principles, models and other architectural prescriptions turn out to be open to multiple interpretations.

## 2. Example of a slightly non-linear relationship (between EA Approach and Architectural Insight)

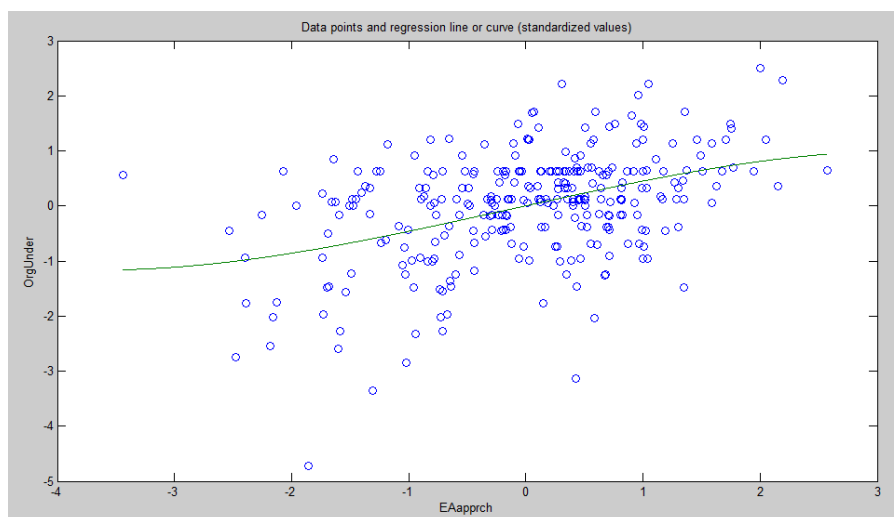

# APPENDIX B: OVERVIEW OF INDIVIDUAL CONTRIBUTIONS

Formative constructs allow summarizing the individual indicators in order to focus on the substantive theoretical relationships at a higher level of abstraction. However, since each formative indicator represents a different aspect, it is advisable to also drill down and study the relationships at the indicator level (Carver 1989). The discussion of individual practices and benefits in Sections 4.3.3 and 4.3.4 is informed by the tables below. These tables provide insight into which individual formative indicators of independent constructs contribute to one or more indicators of dependent constructs. This gives further support for the causal effects in our study and to the composition of constructs. To the best of our knowledge, no methodological instructions exist for this at the time of our research (note that the construct weights cannot be used in our study). We therefore took a critical and conservative approach for verifying the individual contributions: each table column represents a separate PLS model with one dependent variable, with the partialled coefficients of the independent variables in the rows competing with each other (as opposed to mere correlations, which have a higher likelihood of being statistically significant). Therefore, the more rows a column has, the higher the probability of non-significant coefficients. Also note that these columns are not 'final' models, since we did not drop statistically insignificant variables. Rather, they are aimed at providing insight into all variables.

$p<0.1$    $**\ p<0.05$    $***\ p<0.01$    $****\ p<0.001$

**Table C.1: Individual practices of EA Approach**

| EA Approach | Compliance O1 | B2 | B3 | B5 | B9 | B10 | O2 | Organizational Performance B11 |
|---|---|---|---|---|---|---|---|---|
| T3 Managem propagat | 0.235**** | 0.143 | 0.131** | 0.039 | 0.039 | 0.027 | | 0.135** |
| T4 Assess compliance | 0.315**** | 0.001 | 0.065 | 0.118* | | 0.155*** | | 0.199** |
| T7 Assist projects | 0.151** | 0.083 | 0.048 | 0.065 | 0.113** | 0.075 | | 0.187*** |
| T6 Knowledge exch prj | 0.011 | 0.033 | 0.102 | 0.183** | 0.081* | | | |
| T1 Formal approval | -0.120** | -0.060 | -0.089 | -0.015 | -0.104** | -0.051 | | |
| T2 Linked to bus goals | -0.052 | 0.151** | 0.092 | 0.095 | 0.158** | | | 0.165*** |
| T5 Knowledge exch ar | 0.033 | 0.093 | 0.214*** | 0.181*** | 0.058 | | | 0.057 |
| T8 PSA | -0.010 | 0.070 | 0.043 | -0.025 | -0.174** | | | -0.005 |
| T9 Document template | 0.027 | -0.039 | 0.075 | -0.039 | -0.020 | | | 0.024 |
| T10 Financial sanction | 0.043 | -0.048 | -0.013 | 0.040 | -0.037 | | | -0.013 |

**Table C.2: Project Compliance with EA**

| | Compliance O1 | B3 | B9 | O2 | B10 | B2 | B4 | B5 | B8 | B11 |
|---|---|---|---|---|---|---|---|---|---|---|
| O1 Proj compliance | 0.255**** | 0.254**** | 0.290**** | 0.341**** | 0.346**** | 0.226**** | 0.373**** | 0.201** | 0.428**** | 0.197**** |

**Table C.3: Individual Items of Architectural Insight**

| Architectural Insight | B2 | B3 | B5 | B9 | B8 | B10 | B12 | B13 | B14 | B15 | B16 | B17 | B18 | B11 |
|---|---|---|---|---|---|---|---|---|---|---|---|---|---|---|
| B3 Underst complexity | 0.308**** | 0.241** | 0.178** | 0.173** | 0.177** | 0.174** | 0.100 | 0.165 | 0.200** | 0.286**** | 0.217** | 0.181** | 0.130* | 0.212** |
| B9 Clear image future | 0.153** | 0.102 | 0.183** | 0.216*** | 0.148** | 0.093 | 0.050 | 0.022 | -0.006 | 0.093 | 0.041 | -0.046 | 0.069 | 0.233*** |
| B10 Effective communic | 0.222*** | 0.271**** | 0.290**** | 0.289**** | 0.441**** | 0.288**** | 0.077 | 0.019 | 0.163** | 0.174** | 0.191** | 0.069 | 0.279**** | |
| O2 Shared interpretation | 0.116* | 0.047 | 0.041 | 0.051 | 0.137** | 0.158** | 0.164**** | 0.182** | 0.232**** | 0.104 | 0.123** | 0.208**** | 0.193** | 0.164** |

**Table C.4: Individual Items of EA-Induced Capabilities**

| EA-Induced Capabilities | Organizational Performance B1 | B6 | B7 | Project Performance B12 | B13 | B14 | B15 | B16 | B17 | B18 | B11 |
|---|---|---|---|---|---|---|---|---|---|---|---|
| B2 BI/IT alignment | 0.443**** | 0.151** | 0.271**** | 0.098* | 0.007 | 0.089* | 0.173** | 0.133** | 0.097* | 0.108** | 0.373**** |
| B4 Control Complexity | 0.121** | 0.117** | 0.222**** | 0.003 | 0.002 | 0.165** | 0.128 | -0.073 | 0.223*** | 0.125** | 0.164**** |
| B5 Standardize, integr | 0.180*** | 0.308**** | 0.106* | 0.103 | -0.027 | -0.083 | -0.016 | 0.171** | -0.044 | 0.025 | 0.222**** |
| B8 Co-operate | 0.047* | 0.071 | 0.103 | 0.094 | 0.188** | 0.132** | 0.068 | 0.190*** | -0.072 | 0.013 | 0.088* |

**Table C.5: Individual Items of Project Performance**

| Project Performance | Organizational Performance B1 | B6 | B7 | B11 |
|---|---|---|---|---|
| B12 Budgets | 0.053 | 0.121 | 0.057 | 0.004 |
| B13 Deadlines | 0.088 | 0.094 | 0.074 | -0.035 |
| B14 Risk management | 0.056 | 0.219** | 0.013 | 0.035 |
| B15 Quality | 0.123** | 0.015 | 0.004 | 0.101 |
| B16 Functionality | 0.105* | 0.098 | 0.184** | 0.111 |
| B17 Complexity | 0.098 | 0.085 | 0.278**** | 0.275**** |
| B18 Initialization speed | 0.161** | 0.083 | 0.097 | 0.117** |

# Author biographies

## Ralph Foorthuis

UWV, CIO Office and Data Services
La Guardiaweg 116 1040 HG Amsterdam, The Netherlands
ralph.foorthuis@uwv.nl
www.foorthuis.nl

Dr. Ralph M. Foorthuis works as an enterprise architect at UWV. In his professional work he focuses on data registers, process and system integration, web portals, data quality, security, information policy, and systems development. Ralph studied at the University of Amsterdam, where he obtained master degrees in both Communication Science and Informatics. He holds a Ph.D. in Information Systems from Utrecht University. His academic research interests include best practices and models for projects conforming to enterprise architecture (EA), project architecture, EA benefits and drawbacks in practice, assessing compliance with EA, and compliance in general. He has experience with qualitative research and statistical multivariate analysis.

## Marlies van Steenbergen

Sogeti Netherlands
Postbus 76 4130 EB Vianen, The Netherlands
marlies.van.steenbergen@sogeti.nl

Dr. Marlies van Steenbergen is principal consultant of enterprise architecture at Sogeti Netherlands with over fifteen years of experience as an enterprise architect. She obtained her master in Linguistics at Groningen University and in Computer Science at Delft University. She now works as an advisor and a coach in the field of enterprise architecture. Marlies van Steenbergen is one of the founders of DYA®, an approach to developing and improving an effective architectural practice, and co-author of the books "Dynamic Enterprise Architecture, How to Make it Work" and "Building an Enterprise Architecture Practice". She holds a Ph.D. in Information Systems. Her research focuses on the effectiveness of enterprise architecture practices.

## Sjaak Brinkkemper

Utrecht University, Information and Computing Sciences
Princetonplein 5 3584 CH Utrecht, The Netherlands
s.brinkkemper@uu.nl

Dr. Sjaak Brinkkemper is professor of Organization and Information at the Institute of Information and Computing Sciences of Utrecht University in the Netherlands. Before this he was a consultant at the Vanenburg Group and a Chief Architect at Baan. Before Baan he held academic positions at the University of Twente and the University of Nijmegen, both in the Netherlands. He holds a B.Sc. of the University of Amsterdam, an M.Sc. and a Ph.D. in Computer Science of the University of Nijmegen. He has published seven books and more than a hundred papers on his research interests: product software, information systems methodology, meta-modelling, and method engineering.

## Wiel Bruls

IBM Netherlands
David Ricardostraat 2-4
1066 JS Amsterdam, The Netherlands
wiel_bruls@nl.ibm.com

Dr. Wiel A. G. Bruls is an Enterprise architect in IBM's global services business in the Netherlands. He received a Ph.D. in Physics from the University of Utrecht in 1984 after which he joined IBM and worked as lead architect on development of commercially available IT and business systems. In 1996 he joined IBM Services where today he consults with clients on enterprise architecture. Wiel Bruls works as a guest researcher at the University of Utrecht in the field of enterprise architecture. His research interests include the use of enterprise architecture in business/IT alignment, strategic change and associated major business IT transformations.